\documentclass[dvips]{article}
\usepackage{icrc2011}

\def\posttrialssig{$>7.5\sigma$}
\def\excesscounts{$268 \pm 34$}

\def\fittedcentroidsexg{R.A. $20^{h} 20^{m} 00.0^{s}$, Decl. $\mathrm +40^{\circ} 49^{\prime} 12^{\prime\prime}$ (J2000))}
\def\fittedextension{$0.18^{\circ} \pm {0.03^{\circ}}_{stat} \pm {0.02^{\circ}}_{sys}$}

\def\integralflux1tev{ $\rm 1.7 \pm {0.3}_{stat} \pm {--}_{sys} ph^{-1} cm^{-2} s^{-1}$}
\def\integralflux{$\rm 6.9 \times 10^{-12} \ ph^{-1} \ cm^{-2} \ s^{-1}$}

\def\fluxpercentage1tev{$\sim 9.5\%$}

\def\fluxpercentage{5\%}

\def\centroidpulsaroffset{$\sim 0.5^{\circ}$}

\def\snr{SNR~G78.2+2.1}
\def\ver{VER~J2019+407}

\def\fgl{1FGL~J2020.0+4049}
\def\fglpsr{1FGL~J2021.5+4026}
\def\psr{PSR~J2021+4026}

\newcommand\veritas{{\it VERITAS}}

\newcommand\fermi{{\it Fermi}}

\title{Recent Observations of Supernova Remnants with VERITAS}

\newcommand{\etal}{\MakeLowercase{\textit{et al. }}} 
\shorttitle{Weinstein \etal Supernova Remnants with VERITAS}

\authors{A. Weinstein$^{1}$, for the VERITAS Collaboration}
\afiliations{$^1$ Iowa State University}
\email{amandajw@iastate.edu}

\abstract{
Supernova remnants (SNRs) are widely considered the most likely source of cosmic rays
below the knee ($10^{15}$ eV).  Studies of GeV and TeV gamma-ray emission in the vicinity of SNRs, in combination with multi-wavelength observations, can trace and constrain the nature of the charged particle population believed to be accelerated within SNR shocks.  They  may also speak to the diffusion and
propagation of these energetic particles and to the nature of the acceleration mechanisms involved.  We report here on recent observations of SNRs with VERITAS, including the discoveries of VHE gamma-ray emission from from G120.1+1.4 (Tycho's SNR) and from the northwest shell of G78.2+2.1 (gamma-ray source VER
J2019+407, which was discovered as a consequence of the VERITAS Cygnus region survey). }

\begin{document}
\maketitle


\section{Introduction}

Cosmic rays with energies up to $10^{15}$ eV are believed to be accelerated at strong shocks of supernova remnants (SNRs).
Current space-borne (Fermi) and ground-based (VERITAS, HESS, MAGIC) gamma-ray telescopes allow us to study cosmic-ray accelerators via secondary gamma-ray production, either via the inverse Compton process (electrons) or via the decay of neutral pions ($\pi^{0} \rightarrow \gamma\gamma$) produced when high-energy cosmic-ray ions collide inelastically with the ambient medium.  Studies of GeV and TeV gamma-ray emission in the vicinity of SNRs, in combination with multi-wavelength observations, can trace and constrain the nature of the charged particle population believed to be accelerated within SNR shocks.  They may also speak to the diffusion and propagation of these energetic particles and to the nature of the acceleration mechanisms involved.  We discuss here recent detections of two TeV gamma-ray sources by VERITAS, one identified with Tycho's SNR, the other potentially identified with SNR G78.2+2.1, that have the potential to further illuminate this long-standing question.

\section{VERITAS}

The VERITAS instrument is an array of four 12-m imaging atmospheric Cherenkov telescopes located at the Fred Lawrence
Whipple Observatory in southern Arizona.   Each telescope is equipped with a $3.5^{\circ}$ field-of-view (FOV).
The instrument has good (15-25\%) energy resolution over an energy range of 100 GeV - 30 TeV and
excellent angular resolution ($\rm R_{68} < 0.1^{\circ}$), and achieves a sensitivity of $1\%$ of the Crab
Nebula flux in approximately 25 hours.  For more details on the VERITAS instrument and analysis techniques,
please refer to \cite{Holder2011, Holder2006}.

\section{Tycho's SNR}

The SNR, G120.1+1.4, also known as Tycho's SNR,
is the relic of a Type 1a supernova\cite{krause} observed in 1572.  Tycho's SNR is young among Galactic SNRs (438 years).  It shows a distinct shell-like morphology in radio\cite{Dickel} and strong
non-thermal X-ray emission concentrated in the SNR rim, with filaments that have been taken as evidence for electron acceleration\cite{hwang,Bamba,Warren,Katsuda}.

VERITAS observations performed between 2008 and 2010 reveal one of the weakest sources yet detected in TeV gamma rays, with an integral flux above 1 TeV of $\rm 1.87 \pm 0.51_{stat} \times 10^{-13} cm^{-2} s^{-1}$, or $\sim 0.9\%$ of the Crab Nebula emission above the same energy\cite{tychoveritas}.  The TeV photon spectrum can be described by a power law: $\rm dN/dE = C(E/3.42 TeV)^{-{\Gamma}}$, with $\rm \Gamma = 1.95 \pm 0.51_{stat} \pm 0.30_{sys}$ and $\rm C = (1.55 \pm 0.43_{stat} \pm  0.47_{sys}) x 10^{-14} cm^{-2} s^{-1} TeV^{-1}$.

\begin{figure}[!t]
  \vspace{5mm}
  \centering
  \includegraphics[width=3.0in]{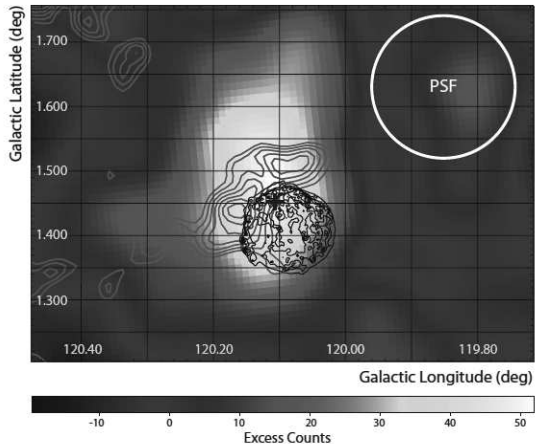}
  \caption{Background-subtracted VERITAS gamma-ray count map showing the region around Tycho's SNR, smoothed with an 0.06° Gaussian kernel. The centroid of the VHE gamma-ray emission is indicated with a thick black cross\cite{tychoveritas}. Overlaid on the image are X-ray contours (thin black lines) from a Chandra ACIS observation\cite{hwang} and 12CO emission (J=1-0) from the FCRAO Survey (gray lines)\cite{heyer1998}.  The VERITAS PSF is shown as a white circle.}
  \label{fig:tychomap}
 \end{figure}

The source is well-fit by a simple symmetric Gaussian with width fixed at the instrument point-spread function ($68\%$ containment radius of $0.11^{\circ}$) and is thus compatible with a point source.  The center of the fit, at 00h 25m 27.0s, +64° 10' 50'' (J2000), is offset by $0.04^{\circ}$ from the center of the remnant, with the statistical and systematic uncertainties in this position being $0.023^{\circ}$ and $0.014^{\circ}$ respectively.  Figure \ref{fig:tychomap} shows
that the peak of the gamma-ray emission is slightly displaced in the direction of a molecular cloud that is seen by
integrating over the velocity range between $\rm \sim 68 km s^{-1}$ and $\rm \sim 50 km s^{-1}$\cite{Lee2004} and that may
be interacting with the northeast quadrant of the remnant.  While it is
tempting to interpret this as indicative of hadronic emission associated with the cloud itself, not only is the effect not statistically significant but the association of the cloud with the remnant has been recently disputed\cite{Tian}.

 \begin{figure}[!t]
  \vspace{5mm}
  \centering
  \includegraphics[width=3.0in]{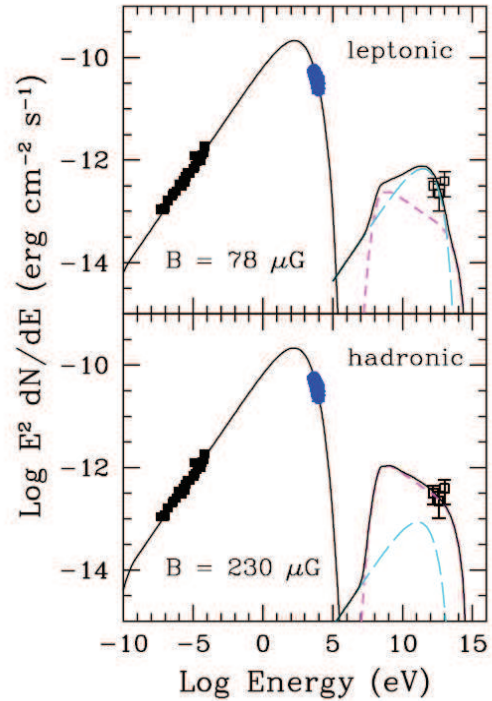}
  \caption{Emission models fitted to the broadband SED (radio, non-thermal X-ray, and VHE gamma-ray emission) of Tycho's SNR. Upper panel and lower panels show a lepton-dominated and hadron-dominated model, respectively. The long-dashed curve is the inverse-Compton (IC) component; the dashed curve the pion-decay emission. The solid line at high energies is the sum of these components while at low energies it corresponds to synchrotron emission.}
  \label{fig:tychomw}
 \end{figure}

Taken together with observations made at other wavelengths, however, the case for hadronic acceleration within the remnant appears strong.  Figure \ref{fig:tychomw} shows a pair of simple model fits, based on scenarios where
the TeV emission is dominated by leptonic (inverse-Compton) emission and hadronic (pion decay) emission\cite{Slane2010} respectively, and generated assuming no influence from the molecular cloud.  Particle spectra of the form
$\rm dN/dE = A E^{-\alpha} e^{(-E/E_{c})}$ have been used, with the spectral index $\alpha$ fixed to a common value
for both scenarios.  The cutoff energy $E_{c}$ is permitted to differ as is expected for loss-dominated
distributions\cite{tychoveritas}.  For the leptonic model the derived field is $\rm \sim 80 {\mu}G$ and the hadronic model
has a lower limit of $\rm \sim 230 {\mu}G$.  These values are well above both the usual $\rm \sim 3 {\mu}G$ fields of the
interstellar medium and the values expected from shock-compression of that field in moderate-density environments\cite{Ellison}.  As a consequence, the VERITAS detection of gamma-rays from Tycho can be considered additional
evidence for magnetic field amplification with the remnant.

With the models as shown here (fitted only to VERITAS, radio, and X-ray data\cite{tychoveritas}, scenarios where the gamma-ray emission is dominated by leptonic or hadronic emission are both a reasonable fit, with the preference for hadrons resting largely on the fact the field of $\rm \sim 80 {\mu}G$ required by the leptonic model is noticeably lower than the
values of $\rm 200-300 {\mu}G$ derived from the thinness of the X-ray synchrotron emission rim\cite{Ballet}.  Along these lines of argument, a more recent paper, incorporating both Fermi-LAT observations of Tycho and the VERITAS spectrum shown here, claims that the only
model consistent with the Tycho broad-band spectrum in this case is one in which the Fermi and VERITAS gamma-ray spectrum is produced by hadrons\cite{Morlino, Naumann-Godo}.  It should be noted, however, that regardless of the model used to explain the gamma-ray emission, the total relativistic particle energy, which represents a significant fraction of the total kinetic energy of the SNR, remains dominated by hadrons\cite{tychoveritas}.

\begin{figure*}[!th]
  \centering
  \includegraphics[width=4.5in]{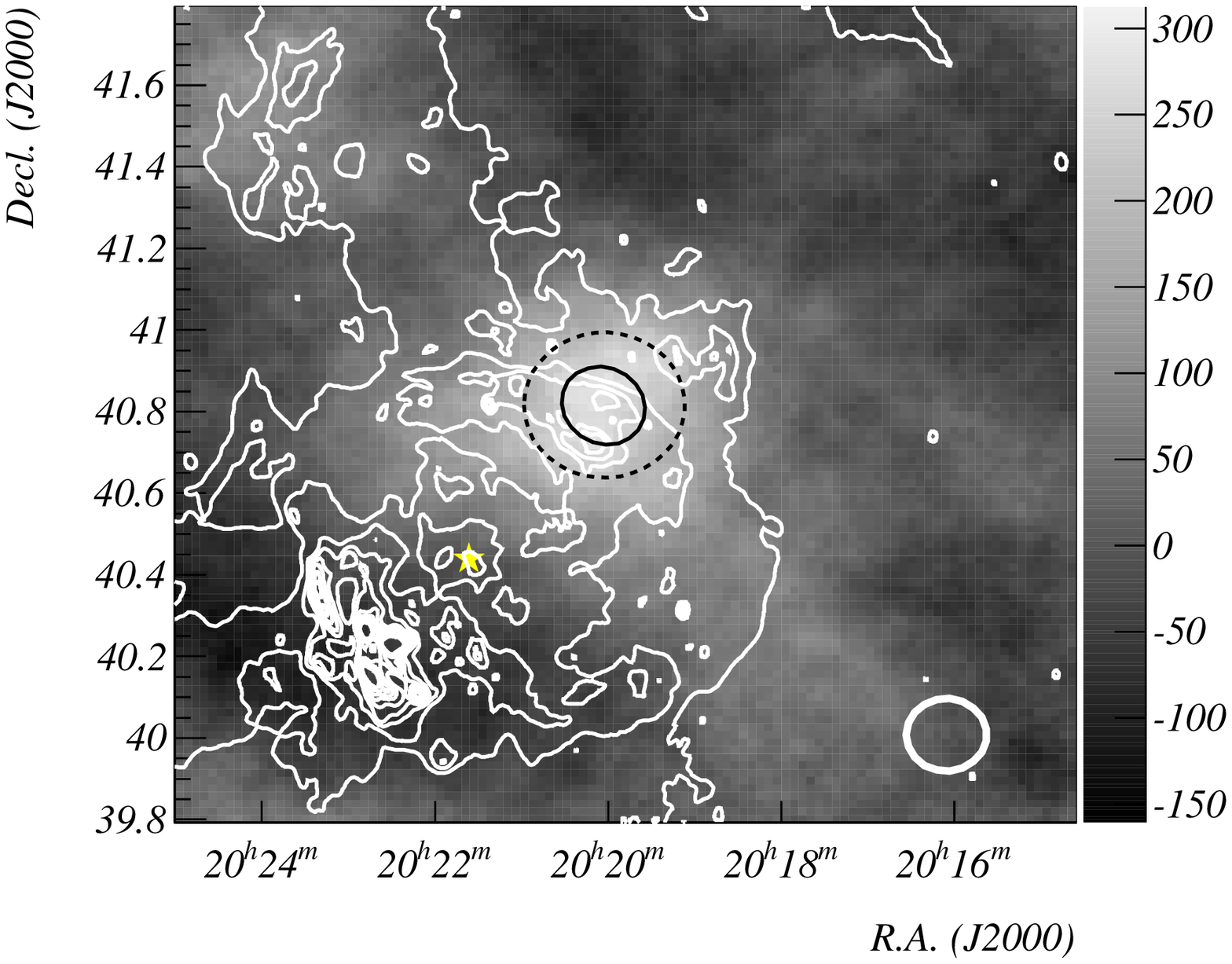}
  \caption{\veritas\ gamma-ray image of \snr\ showing the detection of \ver.
The SNR is delineated by CGPS 1420MHz radio contours (white)
\cite{CGPS};
the solid black ellipse co-incident with \ver\ is the 95\%
confidence ellipse for \fgl; the dashed black circle shows the fitted extent of \ver.  The star symbol shows the location of the
central \fermi\ pulsar \psr.  The white circle (bottom right
corner) indicates the 68\% containment size of the \veritas\ TeV PSF for this analysis.}
  \label{fig:mw}
 \end{figure*}

\section{SNR G78.2+2.1 and VER J2019+407}

SNR~G78.2+2.1 is a $\sim 1^{\circ}$ diameter SNR at $\sim 1.7$~kpc  distance.  At $\sim\!7000$ years \cite{Higgs1977,Landecker1980,Lozinskaya2000} it is considerably older than Tycho, and is thought to be in an early phase of adiabatic expansion into a medium of fairly low density
\cite{Lozinskaya2000}.  Much of the radio and X-ray emission lies in distinct northern and southern
features \cite{Zhang1997,uchiyama}.  Gosachinskij\cite{Gosachinskij2001} also identifies a slowly expanding {\sc H i} shell immediately surrounding the radio shell which Lozinskaya \etal\cite{Lozinskaya2000} suggest was created by the progenitor stellar wind.  A gamma-ray pulsar \psr\ is located at the center of the remnant.  Unlike Tycho, G78.2+2.1's large size makes it possible to resolve in gamma-rays.

The VERITAS Cygnus Region Survey\cite{VERITASCygnusFermi} revealed possible emission in the vicinity of G78.2+2.1, which was confirmed by $\sim 20$ hours of later follow-up observations.  In the follow-up data alone, a clear signal with \excesscounts\ net counts is detected at a location overlapping the northern rim of the remnant, significant at the \posttrialssig\ level after accounting for search trials\cite{verj2019}.  Modeling of the source by a two-dimensional Gaussian convolved with the VERITAS TeV PSF reveals the source to be extended, with an intrinsic extension of \fittedextension and a fitted centroid position of \fittedcentroidsexg.  It should be noted, however, the apparent source profile (as illustrated in Figure \ref{fig:mw}) appears both asymmetric and non-Gaussian.

Figure \ref{fig:mw} shows the gamma-ray excess detected by VERITAS in relationship to the SNR as seen at other wavelengths.  The emission seems to follow an arc-shaped structure in the radio continuum contours (as seen at 1420 MHz).  The brightest part of the emission is offset by \centroidpulsaroffset\ from the \fermi\ LAT pulsar \psr\ (\fglpsr), but is co-located with the Fermi source \fgl\, only $9^{\prime\prime}$ away.  This region of gamma-ray emission does coincide with enhanced thermal X-ray emission that is suggestive of shocked {\sc H i} \cite{uchiyama}; curiously, it lies in a void of CO emission \cite{landp}. The TeV emission also lies near a region of bright
[{\sc S ii}] optical line emission within the SNR that is identified as shock heated gas based on the [{\sc S ii}]/H$\alpha$ line ratio \cite{Mavromatakis2003}.

 \begin{figure}[!t]
  \vspace{5mm}
  \centering
  \includegraphics[width=2.5in]{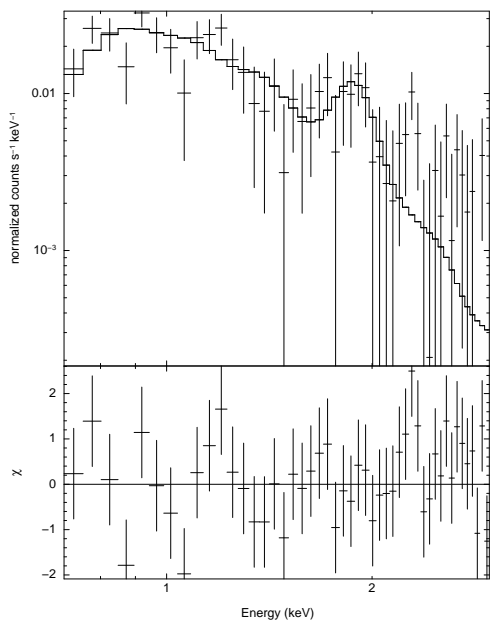}
  \caption{Top: ASCA X-ray spectrum of the region of enhanced X-ray emission coincident with \ver\cite{verj2019}. The stepped line shows the fit of a non-equilibrium plasma model with $\rm kT = 0.5 \pm 0.14 keV$, a column density of $N_H = (3.7 \pm 2.0) \times 10^{21}$ and a normalization of $N = 1.8 \times 10^{-3}$.  The line at 1.7~keV is identified as due to Si; the fit requires an overabundance by a factor of two relative to the solar model.
Bottom Panel --- Residuals from the best-fit model. }
  \label{fig:verj2019_asca}
 \end{figure}

The relationship of \ver\ and \fgl\ to G78.2+2.1, and the nature of the gamma-ray emission being produced,
are not yet clear.  While \ver\ could be a PWN related to \psr, the significant angular displacement between the two sources, when coupled with the total absence of TeV emission at the position of \psr, makes this less likely.
Likewise, while it is possible that \ver\ is the PWN of an unknown pulsar in the line-of-sight towards \snr, this would ascribe the location of \ver\ near the [{\sc S ii}] line emission and the enhanced thermal X-ray emission to chance superposition.  A more straightforward explanation of the TeV emission and the
features observed in the X-ray, optical, and radio continuum would be that the gamma-ray emission arises from particle acceleration (either hadronic or leptonic) within the SNR shock.  High energy electrons capable of producing TeV photons via inverse-Compton scattering should also produce X-ray synchrotron radiation detectable
as a non-thermal power-law in the X-ray spectrum, for which we currently see no evidence in our analysis of the ASCA X-ray spectrum, as shown in Figure \ref{fig:verj2019_asca}.  While molecular clouds are frequently invoked as potential sites of hadronic gamma-ray emission, the dearth of CO in this region means this is not the case here; the target material would have to be {\sc H i}, for which the {\sc H i} shell surrounding the remnant could provide a plausible source.  In this scenario, the pre-shock density estimates of $1-10$~cm$^{-3}$ based on the ratio of [{\sc S ii}] $\lambda$6716 to [{\sc S ii}] $\lambda$6731 line fluxes\cite{Mavromatakis2003} appear consistent with those extrapolated from the TeV gamma-ray flux.

\section{Conclusions}

Detection of gamma-ray emission from Tycho's SNR and from the vicinity of G78.2+2.1 has added to the VERITAS catalog a pair of gamma-ray sources with a strong potential to shed light on cosmic-ray origin and its relationship to supernova remnants.  Although the relationship between the point-like gamma-ray emission from Tycho's SNR and a nearby molecular cloud is not clear, the broad-band SED from this remnant favors not only a predominance of accelerated ions within the remnant but a hadronic origin to the gamma-ray emission itself.  The extended gamma-ray emission from VER J2019+407, on the other hand, appears to track part of G78.2+2.1's shell, which because of the remnant's large size is easily resolved.  While the total lack of CO at this position makes it clear we are not dealing with a molecular cloud scenario, a hadronic origin to the gamma-ray emission from VERJ2019+407 is not ruled out.  The lack of evidence for a non-thermal component at the position of VER J2019+407 would favor hadrons, and preliminary indications are that the density of {\sc H i}, possibly due to interaction with the wall of a cavity blown by the progenitor stellar wind, is sufficient to produce the observed gamma-ray flux in a hadronic emission scenario.

\section{Acknowledgment}

This research is supported by grants from the US Department of Energy, the US National Science Foundation, and the
Smithsonian Institution, by NSERC in Canada, by Science Foundation Ireland, and by STFC in the UK. We acknowledge the
excellent work of the technical support staff at the FLWO and at the collaborating institutions in the construction and
operation of the instrument.

\clearpage

\end{document}